# Theoretical insight into the enhancement of longer-wavelength light absorption in silicon solar cell with multilevel impurities

Shaikh Khaled Mostaque, Bipanko Kumar Mondal, and Jaker Hossain[*]

*Solar Energy Laboratory, Department of Electrical and Electronic Engineering, University of Rajshahi, Rajshahi 6205, Bangladesh.*

**Abstract**

In this article, we theoretically demonstrate multilevel impurity photovoltaic effect in an efficient silicon dual-homojunction solar cell that ensures an extended absorption of longer wavelength light. Along with suitable contact work functions (Ni and Ta as anode and cathode, respectively), three impurity energy levels from acceptor type impurities (One from Tl and two from Zn) have been introduced in the energy gap of the absorber layer in the solar cell. The pristine Si solar cell shows a PCE of 25.4% with $J_{SC}$= 37.99 mA/cm$^2$, $V_{OC}$=0.780V and FF=85.76%, respectively. The incorporation of Tl impurity level alone provides a PCE of 33.4%, with $J_{SC}$= 51.56 mA/cm$^2$, $V_{OC}$=0.789 V and FF=82.03%, respectively. The PCE of the solar cell further enhances to 35.4% with a further enhancement of the short circuit current by 3.76 mA/cm$^2$ due to the inclusion of Zn impurity into the optimized structure. This enhancement of the $J_{SC}$ and hence PCE is resulted from the longer wavelength light absorption due to impurity-assisted two-step photon upconversion in the solar cell.

## 1. Introduction

Present-day photovoltaic industry is being governed by solar cells based on silicon in commercial market. In laboratory, monocrystalline and multi-crystalline silicon based cells have achieved 26.7 and 22.3% respective efficiency[1]. According to detailed balance theory, maximum photoconversion efficiency is 30% for Si with a band gap of 1.1 eV[2]. Alongside, a significant loss takes place due to inability of capturing photons below the energy gap[3]. Either to reach or to overcome this SQ limit, different theoretical and experimental approaches have been proposed. Among them, carrier generation via inclusion of impurity levels, commonly known as impurity photovoltaics (IPV) has already gained popularity. The maximum possibility of photoconversion efficiency with IPV has been figured out theoretically to be 77.2%[4].

In IPV research, the prime objective is to make proper absorption of energies less than the band gap. Establishing one or more impurity energy levels between valance and conduction band allows additional absorption of sub-band gap photons which were previously being wasted in conventional cells. Here, impurity levels provide photons a chance of being absorbed at corresponding energies at sub gaps and accordingly contributing to creation of electron-hole pairs. This in turn plays a role in enhancement of short circuit current $J_{SC}$ and longer wavelength spectral response. The location of the impurity traps between valance and conduction band of host semiconducting material, doping concentration of donor and impurity materials and compensation of charges caused by impurity traps are important specification for significant IPV effect to occur[5]. The impurity levels should be out of the way from the middle of the gap in order to suppress electron-hole recombination through the defects [6]. However, introduced impurity levels also behave as recombination centers that causes decay of open circuit voltage, $V_{OC}$ and overall efficiency[3,7,8]. Some research findings highlighted IPV effect dominating over this decay and contributing to short circuit current density and cell performance[9,10]. These contradictions put emphasis in refining cell performance based on carrier thermal capture cross section or capture probabilities[8].

Till now, different approaches have been implemented numerically or experimentally to find out IPV effect in different hosts e.g. Si[6], SiC[5] or GaAs[11]. Among them Si has drawn preference because of its refractive index having 3.42, which makes use of total internal reflection in a way that empowers multiple passage of rays in the path[7]. Several materials like indium[12], thallium[13], sulfur[7] have gained popularity to introduce one or more impurity levels inside host band gap. Alternative impurity materials e.g. copper[10], tellurium[14], manganese[15], vanadium[5], selenium[16] and magnesium[17] have been studied to find IPV effect in different impurity positions. So far, the optimum position for the impurity trap is located at one third level of the band gap.

The effect of In doping in crystalline Si has been experimentally found to enhance the photocurrent with IPV effect[18]. Though the inclusion of In increased the current up to 6 mA, it ended up with reduction of open circuit voltage $V_{oc}$ and efficiency because of two reasons, (i) depletion of effective doping at base junction when impurity concentration approaches the donor concentration and (ii) decrement of free carrier concentration at base. For further improvement, p-n-n+ structure was proposed with indium concentration of $1 \times 10^{17} cm^{-3}$ where the cell numerically resulted 49.9 mA/cm² of short circuit current[19]. However, indium provides single impurity level at 0.157 eV above valance band[20] whereas 0.20-0.25 eV energy has been found for maximum efficiency[21]. Therefore, thallium with energy gap between 0.24-0.260 eV can be a suitable option for replacing In and enhancing output[13,20,22]. With proper light trapping in the p-n-n+ structure, the short circuit current density reaches around 48 mA/cm² at same donor and impurity concentration of $1 \times 10^{17} cm^{-3}$ with an increment of 9 mA/cm² than with no trapping[13]. It did not cause significant drop in open circuit voltage as well. A comparison between Tl and In and found better improvement with acceptor Tl impurity. However, introduction of impurities from In and Tl together did not come up with significant output[23]. In the same fashion, studies have been further performed with numerical simulations for IPV effect where impurity levels are close to conduction band[7]. Sulfur provides three impurity energy levels in Si at 0.18, 0.37 and 0.52 eV below the conduction band[24]. Being at the middle of the band gap, level at 0.52 eV degraded cell performance. The highest short circuit ($\approx 49.45\ mA$) current density was found for impurity level at 0.37 eV[7]. However, all the energy levels physically are not possible to be appeared at the same time and depend on initializing substance and diffusion conditions[3].

The above studies reflect that the concept of using more than one energy level is not new but extensive researches are yet to be carried out. Few recent reports have been found to enhance photovoltaic parameters putting emphasis either on electron/hole capture cross section or contact work function and back surface field [5,25-27]. To the best of our knowledge, any work has not been published in IPV performance to make use of energy levels above and below mid gap with different impurity materials at a time. To move with this concept, for the first time, we introduce zinc which possess characteristics of sensitizing center for photoconductivity in silicon[28,29]. Studies reveal that zinc alone can provide two acceptor impurity levels at 0.326 and 0.664 eV above the valance band in silicon[30,31].

In this work, studies have been performed adding both Tl and Zn to create multiple impurity traps in silicon absorber with a view to ensure effective absorption of photons at higher wavelengths of solar spectrum with a $p^+$-$n$-$n^+$ structure. So far, previous works have been found to give directions considering a flat band contacts in Si solar cells with IPV effect. Here, we also examined the effect of contact work functions in the solar cell which reveal that suitable contact work functions can also play an integral role in the photovoltaic performance of silicon IPV solar cells.

## 2. IPV device theory and modeling

*2.1 IPV device theory*

In this section, we shortly review the IPV theory. The mechanism of introducing energy level in the band gap is illustrated with the Schokley-Read-Hall (SRH) theory where electron and hole recombination takes place through the traps with thermal excitation[32,33]. To include the IPV effect with conventional SRH model, Keevers and Green proposed a modification where electron and hole transition through impurity considers optical excitation of carriers as well[9].

The Solar Cell Capacitance Simulator (SCAPS), a solar cell device simulator has been found to agree with such modifications and performing simulations with IPV models[10].

The net recombination rate, U via the impurity energy level is given by[9]

$$U = \frac{np - (n_1 + \tau_{n0}g_{nt})(p_1 + \tau_{p0}g_{pt})}{\tau_{n0}(p + p_1 + \tau_{p0}g_{pt}) + \tau_{p0}(n_1 + \tau_{n0}g_{nt})} \quad (1)$$

where, $n_1$ and $p_1$ represent electron and hole concentrations according to

$$n_1 = g_t N_c e^{-(E_c - E_t)/kT} \text{ and } p_1 = \frac{1}{g_t} N_v e^{-(E_t - E_v)/kT} \quad (2)$$

Here, $g_t$ denotes the impurity-level degeneracy factor, $E_t$ is the impurity energy level, $E_v$ and $E_c$ represent valance and conduction band edges, and $N_v$ and $N_c$ are effective densities of states in valance and conduction band, respectively.

In equation (1), $\tau_{n0}$ and $\tau_{p0}$ are low-injection life times for electron and holes from conventional SRH model as

$$\tau_{n0} = 1/c_n N_t \text{ and } \tau_{p0} = 1/c_p N_t \quad (3)$$

where, $c_n$ and $c_p$ are capture coefficients for electron and hole, respectively and $N_t$ represents the defect density.

The terms that describe the IPV phenomena are the optical emission rate of electron and holes from the impurity namely $g_{nt}$ and $g_{pt}$ as

$$g_{nt} = N_t \int_{\lambda_{n\ min}}^{\lambda_{n\ max}} 2\sigma_n^{opt}(x, \lambda) \phi_{ph}(x, \lambda) d\lambda \quad (4)$$

And,

$$g_{pt} = N_t \int_{\lambda_{p\ min}}^{\lambda_{p\ max}} 2\sigma_p^{opt}(x,\lambda)\phi_{ph}(x,\lambda)d\lambda \qquad (5)$$

$\sigma_n^{opt}$ and $\sigma_p^{opt}$ are the electron and hole photoemission cross section of the impurity with threshold wavelengths.

For photon wavelength, $\phi_{ph}(x,\lambda)$ represents the photon flux density at depth $x$ from irradiated window surface.

$$\phi_{ph}(x,\lambda) = \phi(\lambda)\frac{1 + R_b\ e^{-4\alpha_{tot}(\lambda)(L-x)}}{1 - R_f R_b\ e^{-4\alpha_{tot}(\lambda)L}}\ e^{-2\alpha_{tot}(\lambda)x} \qquad (6)$$

Here, $R_f$ and $R_b$ symbolize respective front and back internal reflection coefficients that harmonize to trap corresponding wavelength of light. $L$ is the gross length of the cell structure and $\phi(\lambda)$ is the external AM 1.5G spectrum. $\alpha_{tot}$ represents the absorption coefficient which contains all of the optical mechanisms for absorption.

$$\alpha_{tot}(\lambda) = \alpha_{e-h}(\lambda) + \alpha_n(\lambda) + \alpha_p(\lambda) + \alpha_{fc}(\lambda) \qquad (7)$$

$\alpha_{e-h}(\lambda)$ is the coefficient for inherent generation of electron-hole pairs, $\alpha_n(\lambda)$ and $\alpha_p(\lambda)$ are electron and hole emission coefficients for IPV effect and $\alpha_{fc}(\lambda)$ is a coefficient for free carrier absorption where electron hole pair is not generated.

Due to inclusion of single IPV impurity, the rise of short circuit current density can be represented by[9]

$$J_{IPV} = J_{bare\ IPV} + J_{e-h,\ with\ impurity} - J_{e-h,\ no\ impurity} \qquad (8)$$

Here, $J_{bare\ IPV}$ acts for increment of current due to impurity behavior as a modified SRH recombination center and can be expressed with

$$J_{bare\ IPV} = q \int_{\lambda_{n\ min}}^{\lambda_{n\ max}} \frac{\alpha_n(\lambda)}{\alpha_{tot}(\lambda)} \alpha_{LT}(\lambda)\ \phi(\lambda)d\lambda - \frac{qnpL}{\tau_{n0}(p+p_1)} \qquad (9)$$

where, $\alpha_{LT}(\lambda)$ is described as total absorbance for particular wavelength in the cell.

The difference between current densities due to immersion of impurity and without impurity level accounts the involvement of sub-gap photon current to intrinsic valance band to conduction band absorption current. They are represented by

$$J_{e-h,\ with\ impurity} = q \int_{\lambda\ min}^{\lambda\ max} \frac{\alpha_{e-h}(\lambda)}{\alpha_{tot}(\lambda)} \alpha_{LT}(\lambda)\ \phi(\lambda)d\lambda \qquad (10)$$

$$J_{e-h,\ no\ impurity} = q \int_{\lambda\ min}^{\lambda\ max} \frac{\alpha_{e-h}(\lambda)}{\alpha'_{tot}(\lambda)} \alpha'_{LT}(\lambda)\ \phi(\lambda)d\lambda \qquad (11)$$

The prime terms indicate absence of two absorption process with $\alpha_n(\lambda)$ and $\alpha_p(\lambda)$.

*2.2 Device modeling*

The schematic of silicon solar cell with $p^+$-$n$-$n^+$ structure where three impurity traps has been introduced with Tl and Zn acceptor impurities within the n base layer and the corresponding energy band diagram are displayed in Fig. 1(a) and Fig. 1(b), respectively. From Fig. 1(b), it becomes apparent that quasi Fermi levels $E_{Fn}$ and $E_{Fp}$ do not coincide with valance and conduction band edges. While $E_{Fp}$ lies above $E_v$, $E_{Fn}$ displays its existence below $E_c$. Such a behavior leads to the concept of choosing materials having suitable work function to alleviate collection process of holes and electrons at anode and cathode, respectively. In this work, nickel (Ni) with work function 5.2 eV as anode and tantalum (Ta) with work function 4.0 eV as cathode have been proposed.

The energy band diagram of Fig. 1(b) also includes the schematic representation of IPV phenomena. The inclusion of impurity levels in the solar cell functions as a two-step process by each individual level. In the first step, an electron from valance band is trapped at the impurity level by absorbing a photon (e.g. $h\nu_1$, $h\nu_3$ or $h\nu_5$). In the second step, the trapped electron is pumped to the conduction band by the availability of sufficient photon energy (e.g. $h\nu_2$, $h\nu_4$ or $h\nu_6$). The pumping of electrons corresponds to absorbed photon energy either from direct incident photons or longer wavelength photons reflected from back surface as well as front window. To ensure proper light trapping, the IPV cell is considered to have top surface with antireflection coating and back surface with suitable reflectivity.

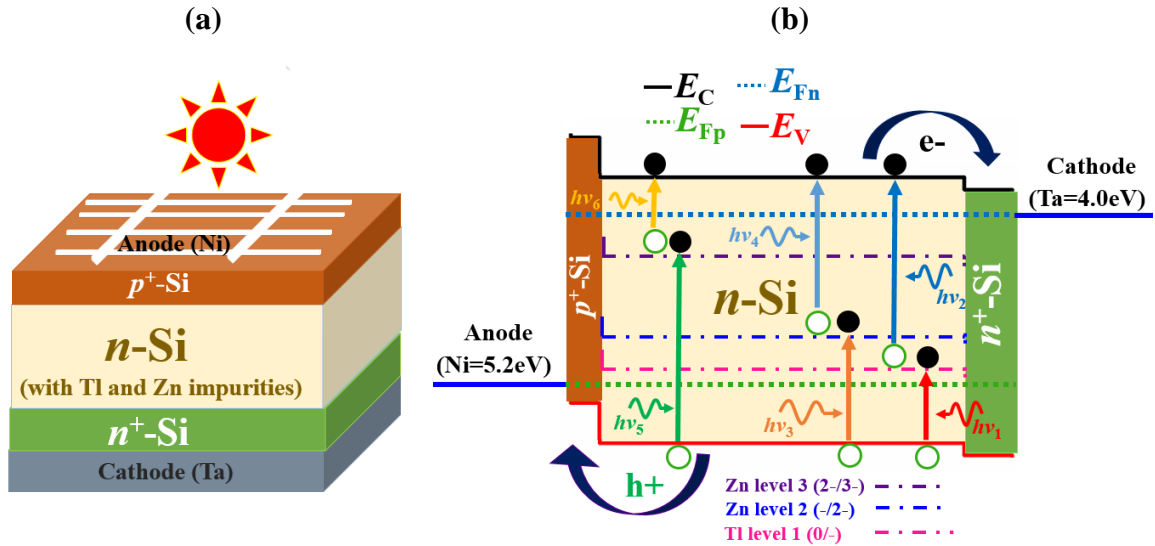

**Figure 1.** (a) The proposed solar cell structure and (b) corresponding energy band diagram with impurity traps showing the impurity-assisted two-step photon upconversion process.

*2.3 Simulation parameters*

The proposed $p^+$-$n$-$n^+$ Si solar cell structure consists of a fine window layer (1-5μm), absorber layer (100-400 μm) and a back surface field (BSF) layer (5-25 μm) with principle parameters of Si shown in Table 1. With insertion of single acceptor impurity Tl, single energy level close to valance band was selected at 0.260 eV above the $E_V$ of Silicon. From previous reports, we set capture cross section of Tl to be $1 \times 10^{-22} cm^2$ and $5 \times 10^{-15} cm^2$, respectively for electrons and holes[23]. The two acceptor levels of Zn were considered at 0.326 and 0.664 eV above the $E_V$ of Silicon. With keeping an analogy with laboratory findings, we varied electron capture cross section from $1 \times 10^{-20}$ to $1 \times 10^{-17} cm^2$ for both energy levels of Zn impurity. Likewise, we examined the effect of the variation of hole capture cross section from $1 \times 10^{-16}$ to $1 \times 10^{-13} cm^2$ [28].

**Table 1:** Simulation parameters for silicon (Si) used in this study.

| **Properties** | *value*[7] |
|---|---|
| Band gap (eV) | 1.120 |
| Electron affinity (eV) | 4.050 |
| Dielectric permittivity (relative) | 11.900 |
| CB effective density of states (1/cm$^3$) | 2.80×10$^{19}$ |
| VB effective density of states (1/cm$^3$) | 1.04×10$^{19}$ |
| Electron thermal velocity (cm/s) | 1×10$^7$ |
| Hole thermal velocity (cm/s) | 1×10$^7$ |
| Electron mobility (cm²/Vs) | 1.35×10$^3$ |
| Hole mobility (cm²/Vs) | 4.80×10$^2$ |
| Shallow window uniform acceptor density, $N_{A,\,window}$ (1/cm$^3$) | $1 \times 10^{18}$ |
| Shallow window uniform donor density, $N_{D,\,window}$ (1/cm$^3$) | 0 |
| Shallow BSF uniform donor density, $N_{D,\,BSF}$ (1/cm$^3$) | $1 \times 10^{18}$ |
| Shallow BSF uniform acceptor density, $N_{A,\,BSF}$ (1/cm$^3$) | 0 |

| Refractive index | 3.42 |
| --- | --- |
| Effective mass of electron | 1.08 |
| Effective mass of hole | 0.55 |

To add up the electron and hole photoemission cross sections of the Tl and Zn, $\sigma_n^{opt}$ and $\sigma_p^{opt}$ have been considered to be zero for photons with energy exceeding the band gap according to models from Lucovsky[34] as well as Keevers and Green[9]. Free carrier absorption has been neglected in the model. To perceive the benefit from IPV effect, characterization of light trapping from front and back of the cell should be ensured[7]. Here, we assumed ideal light trapping considering $R_b$ and $R_f$ equal to unity. Later, to relate with practical consequence, variation of $R_b$ and $R_f$ has been observed.

The proposed cell performance has been judged separately in the presence and in the absence of impurities under standard illumination (AM1.5G spectrum, 1000 W/m² of incident light power and 300 K temperature). The effects of varying thickness, doping concentration and impurity concentration have been investigated with adding individual impurities. Finally, we explored whether any change in window or BSF layer can make any effective contribution to the proposed structure.

## 3. Results and discussion

### 3.1 IPV effect of Tl single impurity in Si solar cell

First, we have investigated the pristine and Tl impurity incorporated Si dual-homojunction solar cell under flat and non-flat band conditions, respectively. Then, the effects of Si absorber layer and the Tl impurity concentration on the photovoltaic performance have also been discussed in details.

*3.1.1 Role of contact work functions on PV performance*

Fig. 2 shows J-V characteristics of the proposed silicon solar cell structure with and without Tl impurity under flat and non-flat band condition, respectively at an absorber layer doping concentration of $N_D= 2\times10^{17}$ cm$^{-3}$. As observed in the figure, the pristine Si solar cell without any IPV provides a PCE of 20.1% with Jsc= 34.49 mA/cm$^2$, Voc=0.690 V and FF=84.47% under flat band configuration. The introduction of Ni (WF=5.2 eV) as anode and Ta (WF=4.0 eV) as cathode into the structure not only improves the current density but also maintains a higher open circuit voltage of 0.780 V. The $J_{SC}$ of the device increases to 37.99 mA/cm$^2$ with a PCE of 25.4% and FF of 85.76%, respectively under non-flat band condition. The reasons behind of these improvements of current and voltage can be explained as follows:

When a metal is at contact with the absorber layer, the Schottky barrier, $\phi_s$ at the metal contact is given by $\phi_s = \phi_m - E_V$. Where, $\phi_m$ is the metal work function and $E_V$ is the valence band energy. At the front contact, a Schottky barrier forms if $\phi_m < E_g+(\delta_{Abs}+\phi_d)$ where, $E_g$ is band gap, $\delta_{Abs}$ is electron affinity of absorber material, $\phi_d$ presents the energy difference in the valence (or conduction) band in bulk and at the edge. The bending of the energy band at the back contact occurred owing to the surface defect states and high surface recombination velocities which corresponds to the Fermi level pinning and high Schottky barrier lowers the $V_{bi}$ in a typical device. The ohmic contact is formed when the Schottky barrier height is reduced. For p$^+$ absorber layer, a metal with a high work function ($\phi_m$) close to ($\delta_{Abs}+\phi_d+E_g$) should be chosen to avoid Schottky barrier[27]. A similar case is happened for the n$^+$ BSF layer in the back (cathode) contact. Therefore, the reduction of Schottky barrier in both cathode and anode contacts will enhance the built-in potential which will in turn enhances the $V_{OC}$ and $J_{SC}$ of the devices as they facilitates the easy path for carrier collection[27].

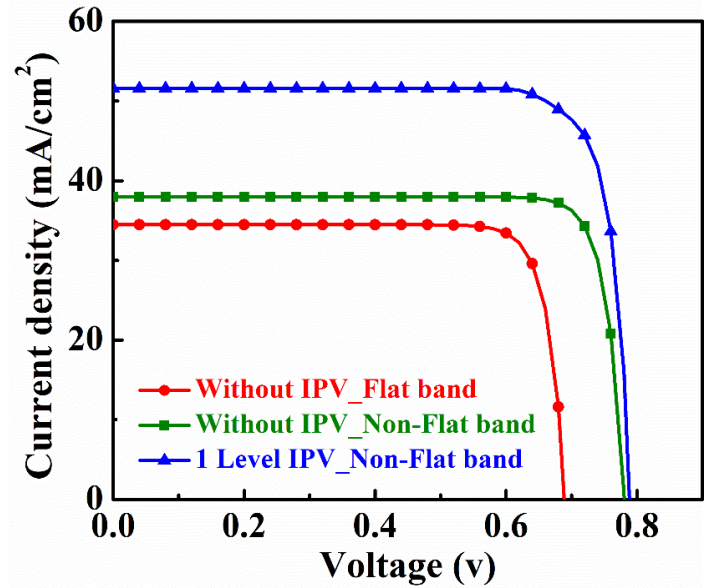

**Figure 2.** The J-V characteristics of Si solar cells with zero IPV and one level of impurity under flat and non-flat band conditions.

However, the insertion of first impurity level (Tl) close to valance band with Tl concentration of $2\times10^{17}$ cm$^{-3}$ in absorber layer further rises current to 51.56 mA/cm$^2$ which clearly exhibits the enhancement of solar cell performance. This enhancement of short circuit current is mainly due to the absorption sub-band gap photons in the absorber layer with impurity.

*3.1.2 Absorber layer effect on PV performance*

One of the most important factors in optimizing solar cell performance is the active layer thickness. Fig. 3a depicts the consequence of varying Si-absorber thickness on PV performance. Shallow uniform donor density, N$_D$ and uniform defect density, N$_t$ of $2\times10^{17}$ and $7\times10^{13}$ cm$^{-3}$, respectively were considered during the simulation. While flat band consideration exhibits a gradual decrement in J$_{SC}$ and PCE with the increment in absorber thickness, non-flat band combination attempts to keep a steady value with gradual refinement in FF as seen in the figure. Though an improved V$_{OC}$ is observed in non-flat band case, the voltages in both cases did not change with absorber thickness and doping concentration. The safeguarding of V$_{OC}$ is caused by the ability of keeping high built-in potential in the chosen structure of $p^+$-$n$-$n^+$[19]. The high built in potential holds occupancy of Tl impurity level. Flat band design holds up open circuit voltage of 0.68 V while non-flat band formation delivers consistent 0.79 V over the range. The $J_{SC}$ decreases with increase in thickness both in flat and non-flat condition and provides the highest values at 100 μm. In this instance, non-flat band structure results with 51.56 mA/cm$^2$ of current density and 82.03% of fill factor while flat band stays below 46.82 mA/cm$^2$ with an FF of 79.22%. Consequently, flat band PCE displays low values comparable to non-flat band and follows a downward trend with respective $J_{SC}$. This can be explained as the conversion efficiency and the short circuit current density is limited by SRH recombination as described in equation 1[35]. Moreover, higher doping concentrations may have an impact on both electron and hole diffusion lengths[36]. On the contrary, though flat band $J_{SC}$ displays a gradual decrement, corresponding FF has an upward trend. Higher values of $J_{SC}$ and FF keep non-flat band PCE smooth with negligible changes from 33.4% over the range.

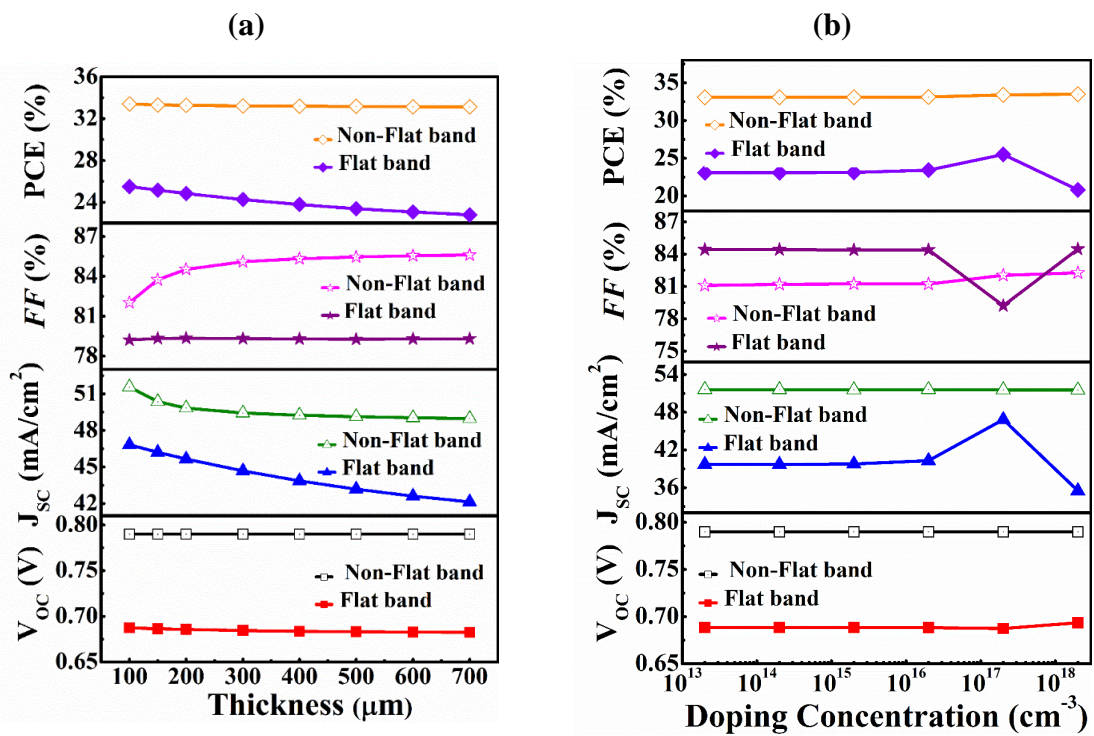

**Figure 3.** The change of PV parameters with (a) absorber layer thickness and (b) doping concentration in Si solar cell having Tl impurity only.

The outcome of changing doping concentration Si absorber layer in Si IPV solar cell with Tl impurity is depicted in Fig. 3b. The figure clearly distinguishes between a flat band and non-flat band configuration for the studied model. The conventional flat band structure maintains open circuit voltage at 0.69 V over a long range of doping concentration taking the advantage of safeguarding $V_{OC}$ by the proposed structure[19]. Though, $J_{SC}$ begins rising to reach a peak value from doping concentration of $2\times10^{16}$ cm$^{-3}$ for the flat band, corresponding FF starts turning down to maintain a steady $V_{OC}$. The $J_{SC}$ reaches its peak value of 46.82 mA/cm$^2$ at a doping concentration of $2\times10^{17}$ cm$^{-3}$ but FF declines to its trough at the same moment resulting a PCE of 25.5% only. This result is consistent with previous work which shows that $J_{SC}$ reaches its peak pick when doping concentration, $N_D$ equals impurity concentration, $N_t$ i.e. in the order of $10^{17}$ cm$^{-3}$ [13,23]. The amount of doping concentration, $N_D$ determines the consignment of free carriers and plays a role in the recombination rate as described in equation 1. As the volume of $N_D$ reaches $N_t$, the slowest optical process illustrated by $\sigma_n^{opt}$ and $\sigma_p^{opt}$ takes part in enhancement of $J_{SC}$. However, higher order of doping concentration also constricts the band gap causing intrinsic band to band absorption to shift to lower energies[13]. The corresponding PCE curve follows the $J_{SC}$ pattern since $V_{OC}$ exhibits a constant behavior. Though both $V_{OC}$ and $J_{SC}$ follows a upward trend over a long range of doping concentration, a drop in fill factor may occur due to rise in series resistance, a decrease in shunt resistance, or a combination of the two[12]. On the other hand, non-flat band structure seems to maintain a higher $V_{OC}$ (0.79V) as well as higher $J_{SC}$ (51.56 mA/cm$^2$) than in flat band configuration. The smooth nature of $V_{OC}$ and $J_{SC}$ provide a steady FF and better PCE, 82.03 and 33.4% respectively, proving the utility of connecting metals with suitable work function at anode and cathode.

*3.1.3 Effect of impurity concentration on PV performance*

Fig. 4 shows the variation of solar cell performances with the concentration of Tl impurity. It is revealed from the figure that the maximum efficiency appears when Tl concentration, $N_t$ is equal to Si concentration, $N_D$ in absorber for flat band condition. Note that, impurity concentration is being varied keeping the absorber $N_D = 2 \times 10^{17} cm^{-3}$. At values $N_t < N_D$, short circuit current density, J$_{SC}$ inflates after $N_t = 1.8 \times 10^{16}$ cm$^{-3}$. Such an increment originates from the absorption of sub-band gap photons which is described by $g_{nt}$ and $\sigma_n^{opt}$ in equation 4. When N$_t$ >N$_D$, the impurity energy level yields some unoccupied states which in turn increases transition rate from valance band to impurity level but decreases the photon flux density $\phi_{ph}(x, \lambda)$ required for emission from impurity level to conduction band. Thus, a rapid decrement observed in J$_{SC}$ and consequently in PCE. In flat band condition, the proposed structure displays analogous performance with previous simulation works[13,23]. However, with non-flat band, numerical performance displays improvement on maximum value of J$_{SC}$ (51.56 mA/cm$^2$) and PCE (33.4%) at lower impurity concentration of $5 \times 10^{12} cm^{-3}$ while in flat band connection maximum J$_{SC}$ resulted 46.82 mA/cm$^2$ with 25.49% PCE at N$_t$=N$_D$. The reason of such slight improvement is demonstrated in section 3.1.1.

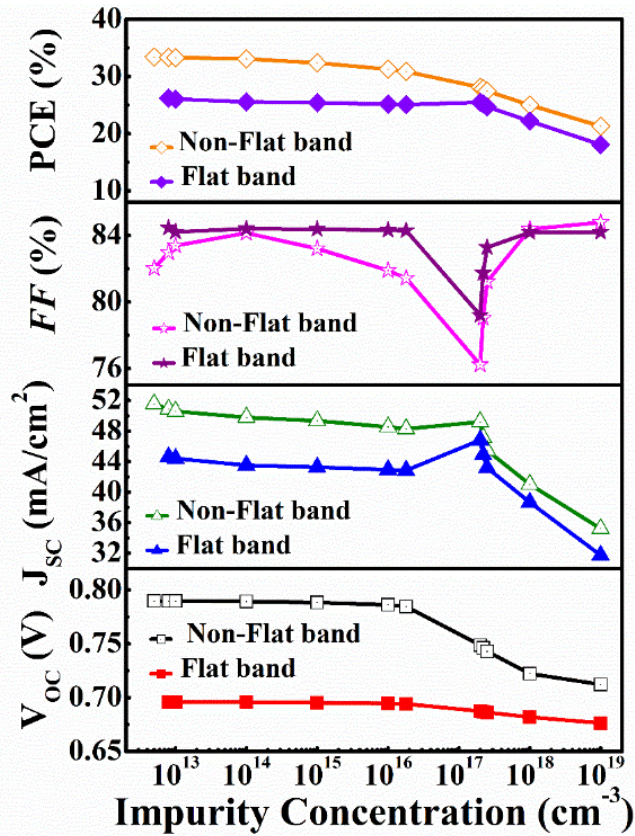

**Figure 4.** The change of PV parameters with impurity concentration of Si solar cell with single Tl impurity.

*3.1.4: Effect of light trapping on PV performance*

The significance of light trapping in IPV solar cell has already been suggested by Keevers and Green[9]. Light trapping enhanced the probability for photons to be absorbed by weak optical processes in a solar cell. It is possible to maximize the optical path and enhance the absorption probability for photons by weak optical process like emission by $h\nu_1$ in Fig. 1(b). To evaluate such effect of light trapping due to longer wavelength photon absorption in single Tl impurity level, we demonstrate J-V characteristics and corresponding spectral response in Fig. 5 and 5b. Here , the degree of light trapping has been regulated by the value of front internal reflection $R_f$ and back surface reflection $R_b$ as demonstrated by Khelifi *et al.*[37]. The maximum possible absorption was obtained by considering a perfect internal reflection at the front as well as reflection from back surface with $R_f=R_b=1$. A notably higher short circuit current resulted due to the absorption of photons with extension of λ upto 1400 nm. While with ideal light trapping, $J_{SC}$ comes up with 51.56 mA/cm$^2$, adjusting the $R_f$ and $R_b$ to 0.9999 results a $J_{SC}$ of 46.65 mA/cm$^2$ and PCE of 31.62% with FF=85.94%. At $R_f=R_b=0$, i.e. for no light trapping, the PV parameters matches with the values that were in pristine structure. The spectral response in this case also reveals that light above 1100 nm has not been absorbed by the solar cell. However, it can be noted that there is already a proposed Bragg-reflector structure consisting of thin alternating layers of AlAs and Al$_x$(Ga$_{1-x}$)As that can nearly reflect 100% of the longer-wavelength photons[11].

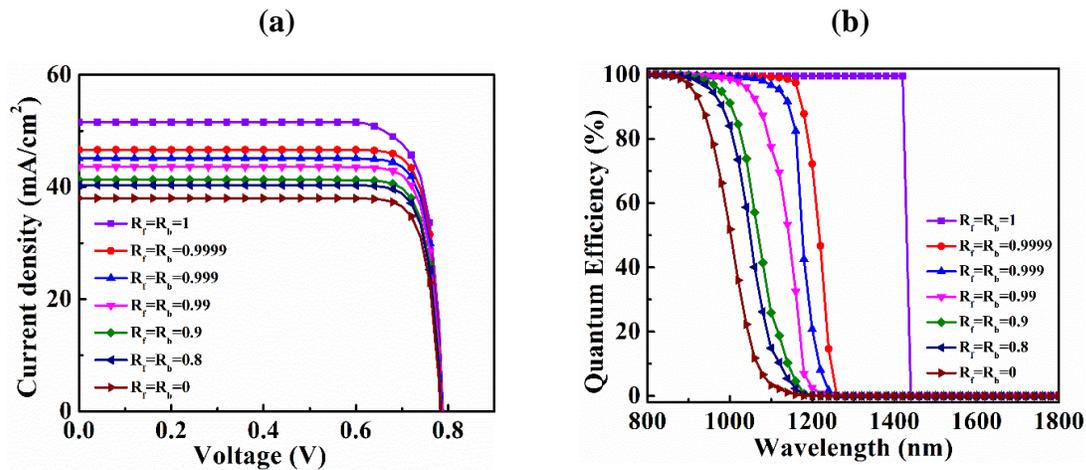

**Figure 5.** Effect of light trapping on single level IPV (a) J-V characteristics and (b) sub-band gap spectral response.

**3.2 IPV effect of multilevel impurities (Tl and Zn) on PV performance of Si solar cell**

In the previous sections, Tl impurity providing energy level close to valance band is improving short circuit current density, $J_{SC}$ and PCE through absorption of sub-band gap photons. Appending Zn impurity levels at 0.326 and 0.664 eV creates further two energy levels below and above the absorber mid gap[30,31]. Naturally, we can expect that the sub-band gap photons should contribute much more current if proper light trapping is established. In this section, the numerical simulation performance of Tl+Zn impurities that provides three impurity levels in silicon absorber has been discussed in details.

The optimized structure considers absorber thickness=100 μm, shallow uniform donor density, $N_D=2\times10^{17}$ cm$^{-3}$ and uniform impurity concentration, $N_t=7\times10^{13}$ cm$^{-3}$. The electron and hole capture cross section for Tl were taken $1\times10^{-22}$ and $5.07\times10^{-15}$ cm$^2$, respectively. For both impurity levels of Zn, corresponding electron and hole capture cross section were $1\times10^{-20}$ and $1\times10^{-16}$ cm$^2$, respectively[28]. The thickness of $p^+$ type window layer and $n^+$ type BSF layer were set to 5μm and 10μm, respectively with analogous acceptor and donor density of $1\times10^{18}$ cm$^{-3}$. It is also noted that front and back reflection was set to unity (that is possible to achieve by Bragg-reflector for the device simulation with Tl and Zn impurities)[11].

*3.2.1. Effect of Si absorber layer on PV performance*

Fig. 6a presents the consequences of increasing thickness of absorber in the solar cell structure with three impurity levels. On optimized value of impurity concentration, $V_{OC}$ does not have any ramifications with gradual addendum of absorber thickness or Si doping concentration in absorber layer. The proposed structure holds up $V_{OC}$ at 0.79 V with trivial change in FF. As discussed in section 3.1.2, the $p^+$-$n$-$n^+$ structure is still responsible for holding up the $V_{OC}$, while the role of correct work function is responsible for further rise of the $V_{OC}$ level[19,20]. On the other hand, the combination of Tl and Zn impurities may further speed up the recombination process and also some defect levels are significantly beyond the optimal impurity level that lower the short circuit current and hence the performance of IPV solar cells[3]. For instance, variation in thickness from 100 to 400 μm, gradual declination of $J_{SC}$, which falls from 55.32 to 45.33 mA/cm$^2$ causes PCE to plummet at 28.5% from 35.4%.

Thus, the optimized thickness of absorber layer is desirable to obtain a higher efficiency with the proposed structure.

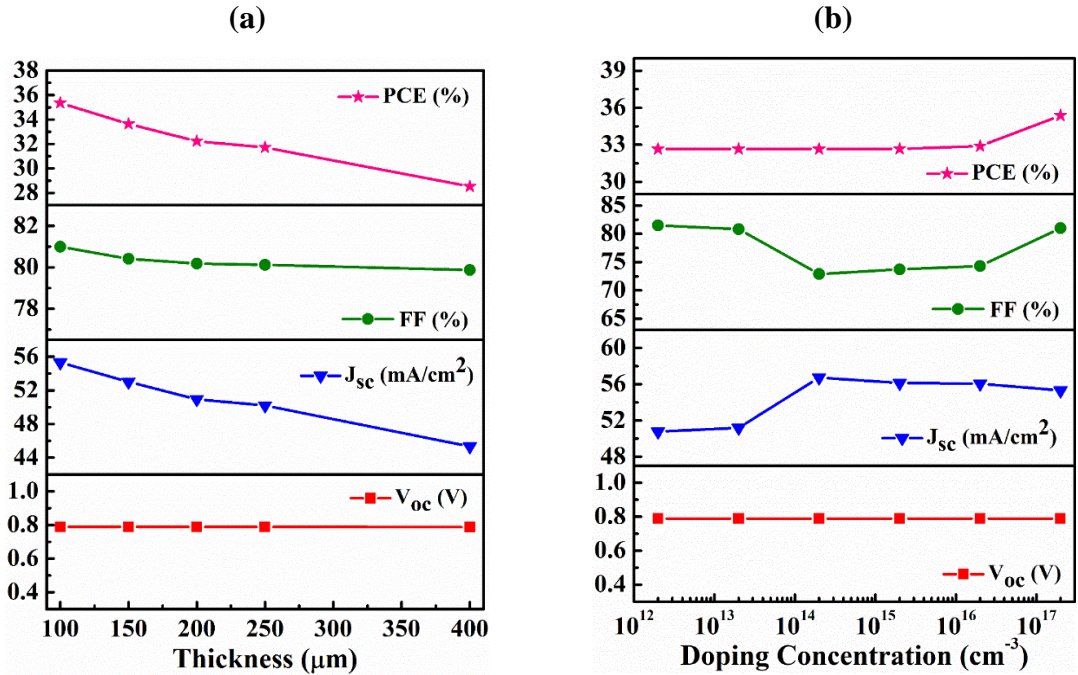

**Figure 6.** The variation of PV parameters with absorber (a) thickness and (b) doping concentration of Si solar cell having multilevel impurities.

Fig. 6b illustrates the advantages and disadvantages of different level of Si doping concentration in the absorber layer of the solar cell structure with three impurity levels. With the $p^+$-$n$-$n^+$ structure advantage, no significant change in $V_{OC}$ (0.79 V) is noticed[19]. The inclusion of both Tl and Zn impurities plays a complex role in the improvement of the other parameters i.e. $J_{SC}$, FF and PCE. In lower orders of doping, $J_{SC}$ is low but FF is high. At the order of $10^{13}$, $J_{SC}$ starts climbing up and reaches peak 56.74 mA/cm$^2$ at a concentration of $2\times10^{14}$ cm$^{-3}$. On the other hand, FF faces a consequent decrease to 72.88% resulting no improvement over PCE. Moreover, with further doping FF starts to rise from the order of $10^{16}$ and PCE reaches at 35.36% with 55.31 mA/cm$^2$ current density in the concentration of $2\times10^{17}$ cm$^{-3}$. The initial drop in fill factor may be attributed to an increase in series resistance, a decrease in shunt resistance, or a combination of both[12]. However, an increased uniform absorber doping concentration (> $10^{14}$ cm$^{-3}$) can counteract the effect of increased or decreased resistance, improving FF and PCE. On the contrary, $J_{SC}$ still exhibits a slight downward movement after certain order of doping concentration with $N_D>N_t$. The inclusion of three separate impurity levels can be responsible for intensifying the recombination process at higher orders of doping[3].

*3.2.2 Effect of multilevel impurity concentration on PV performance*

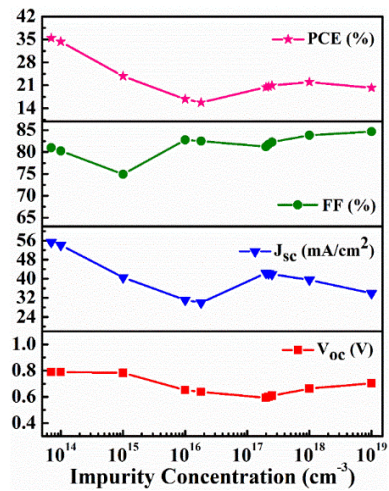

**Figure 7.** The variation of PV parameters with multilevel impurity concentration in Si solar cell.

Fig. 7 reveals the impact of different order of impurity concentration, $N_t$ on PV parameters while keeping the Si doping concentration at $N_D=2\times10^{17}$ cm$^{-3}$ in the absorber layer. Uniform impurity concentration upto the order of $10^{15}$ does not show any significant change in $V_{OC}$. With higher orders, slight variation in $V_{OC}$ is observed. Thus, $V_{OC}$ keeps almost a stable performance indicating the advantage of the solar cell structure. The value of $J_{SC}$ gets a significant drop up to increasing concentration to the order of $10^{16}$. After that, it starts to rise till it reaches another peak of 42.37 mA/cm² at $N_t=N_D$. At $N_t>N_D$, $J_{SC}$ again starts to fall in lack of required the photon flux density $\phi_{ph}(x,\lambda)$ for emission of electron from impurity level to conduction band. Thus, a non-flat band combination becomes an advantage to obtain 55.32 mA/cm² of current density with 35.4% PCE at lower impurity concentration at $N_t = 7 \times 10^{13} cm^{-3}$.

*3.2.3 Window layer effect on PV performance*

In this section, the effect of window layer thickness and doping on the photovoltaic performance has been studied in details.

Fig. 8a shows that no significant variation on the PV parameters is observed due to the change in thickness of the window layer in the range of 1-5 µm. Here, the window layer have been chosen with small thickness to form a proper p-n junction with the absorber as well as to pass most of the spectrum to absorber[38]. However, with extreme doping of p type Si window causes progressive reduction in $J_{SC}$ and corresponding PCE of the solar cell as illustrated in Fig. 8b. For a doping of the order of $10^{19}$ cm⁻³, $J_{SC}$ exhibits exponential decrement and reaches at 41.12 mA/cm² with PCE of 26.12% for a doping of the order of $10^{21}$ cm⁻³. The higher carrier concentration in the window layer may decrease the diffusion length of the charge carriers that results the decrease in $J_{SC}$ of the solar cell[39]. Therefore, thinner window and a doping just above the absorber layer can be useful for forming a proper junction with absorber and allowing visible and infrared part of the light to enter into absorber layer.

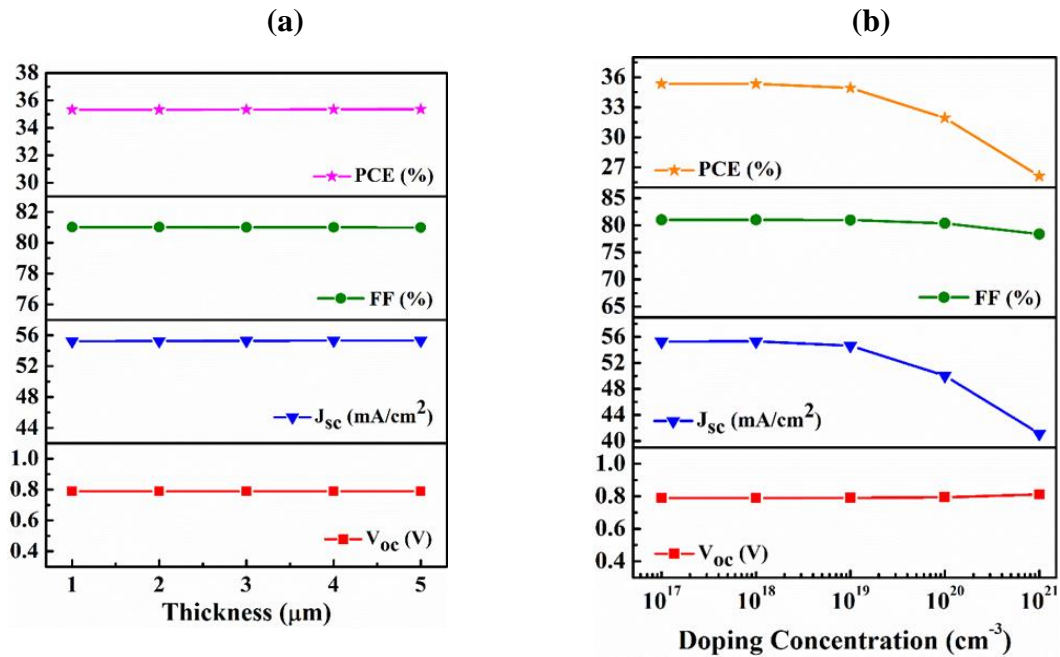

**Figure 8.** Effect of (a) thickness and (b) doping concentration of window layer on PV performance of Si solar cell with multilevel impurities.

*3.2.4 BSF layer effect on PV performance*

This section illustrates the impact of thickness and doping concentration of BSF layer on the performance of Si solar cell with multilevel impurities. The change of thickness and doping concentration of BSF layer results a quite few changes on PV parameters as signified in Fig. 9a and b, respectively. For a 20 μm change in BSF thickness, $J_{SC}$ and PCE exhibit low decrement of 0.8 mA/cm$^2$ and 0.06%, respectively while keeping $V_{OC}$ and FF at constant values. Similarly, with the change in doping concentration, all PV parameters proclaim steady values. A previous study of a Si heterojunction cell that demonstrates PV performance over a wide range of BSF thickness and doping concentration reported a similar response[40]. Thin BSF layer appears to have little effect on solar performance, even when BSF is used to achieve high efficiency[41]. Thus, to build up n-n$^+$ junction between absorber and BSF layer it is desirable to maintain the least possible thin layer with doping above the value of that of the absorber layer.

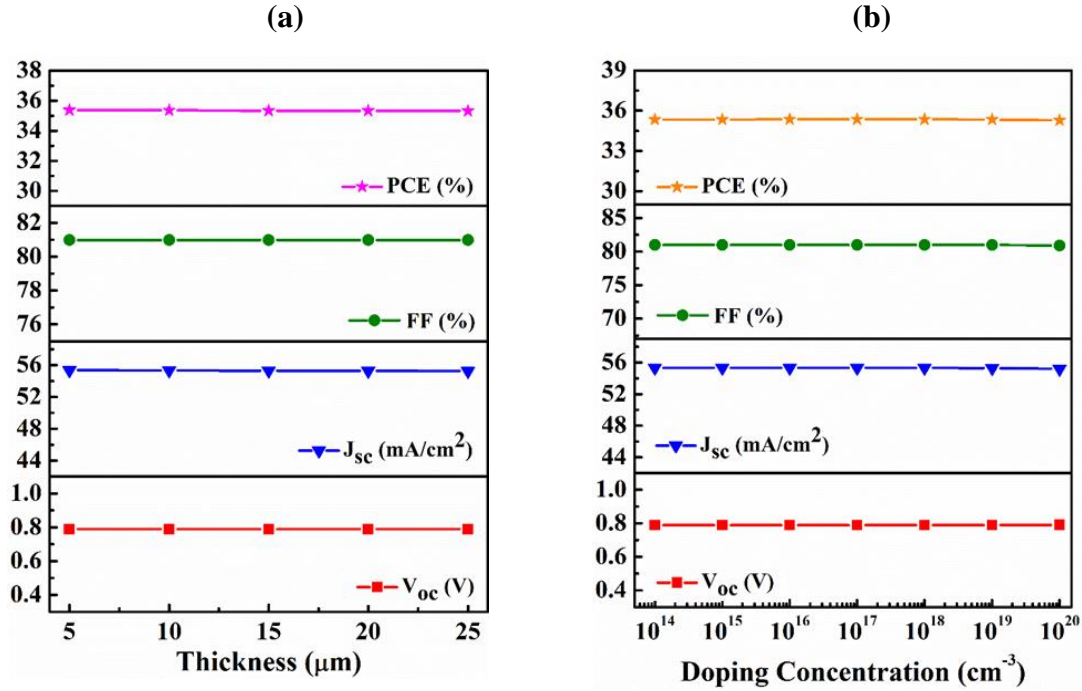

**Figure 9.** The effect of (a) thickness and (b) doping concentration of back surface field (BSF) layer on PV performance of Si solar cell with multilevel impurities.

*3.2.5 Role of thermal cross section on PV performance*

The IPV paradigm relates absorption of photons with energies less than the band gap through reflecting back longer wavelength light from back surface. As described in Fig. 1b, the additional impurity levels inside the band gap provide a platform for absorption of sub-band gap photons and consequently augmentation of short circuit current density, $J_{SC}$. However, the process can lead to the degradation of PV parameters if impurity levels act as recombination centers[42]. Suitable thermal capture cross sections namely electron and hole capture cross sections may provide an indication to make IPV to perform in enhancement of $J_{SC}$ and regulate the degradation of $V_{OC}$[8]. For simulation, we considered absorber doping, $N_D=2\times10^{17}$ cm$^{-3}$ and uniform $N_t=7\times10^{13}$ cm$^{-3}$, respectively. The electron and hole capture cross section for single impurity Tl was kept fixed at $1\times10^{-22}$ and $5.07\times10^{-15}$ cm$^2$, respectively. We considered same thermal cross sections for both energy levels of Zn impurity.

*3.2.5.1 Role of electron and hole capture cross section of Zn*

In this section, the effect of extending capture cross-section area and impurity concentration on the $J_{SC}$ has been depicted in order to obtain an optimized condition for maximum short circuit current of the solar cell. The capture cross-section areas for electron and hole have been varied within the range of $10^{-17}$-$10^{-20}$ and $10^{-13}$-$10^{-16}$ cm$^2$, respectively[28]. Fig. 10a represents the contour plot to the show two dimensional view for change in short circuit current density, $J_{SC}$ of the proposed structure with impurity concentration and electron capture cross section. The electron capture cross section was varied within the range of $10^{-17}$-$10^{-20}$ cm$^2$ while keeping the hole capture cross section at $1\times10^{-16}$ cm$^2$. At cross section of $10^{-20}$ cm$^2$ and impurity concentration of $7\times10^{13}$ cm$^{-3}$, $J_{SC}$ has the highest value. At this level, $J_{SC}$ does not show declination even if the impurity concentration is put in the order of $10^{19}$ cm$^{-3}$, nevertheless $J_{SC}$ decreases beyond this level. The $J_{SC}$ gradually goes down with increasing the order of capture cross section and exhibits a poor value beyond the order of $10^{-18}$ cm$^{-2}$. On the other hand, Fig. 10b represents the variation of short circuit current with impurity and hole capture cross section with a constant electron capture cross section of $1\times10^{-20}$ cm$^2$. The $J_{SC}$ has the higher value at the lower values of hole capture cross section. For an increase capture cross section from $10^{-14}$ to $10^{-13}$ cm$^2$, $J_{SC}$ results with lower values even with impurity concentration in the order of $10^{14}$ cm$^2$. These results resemble a previous work where electron capture cross section greater than $10^{-20}$ cm$^2$ reflects a negative gain and higher order of hole capture cross section creates detrimental performance with IPV effect[5].

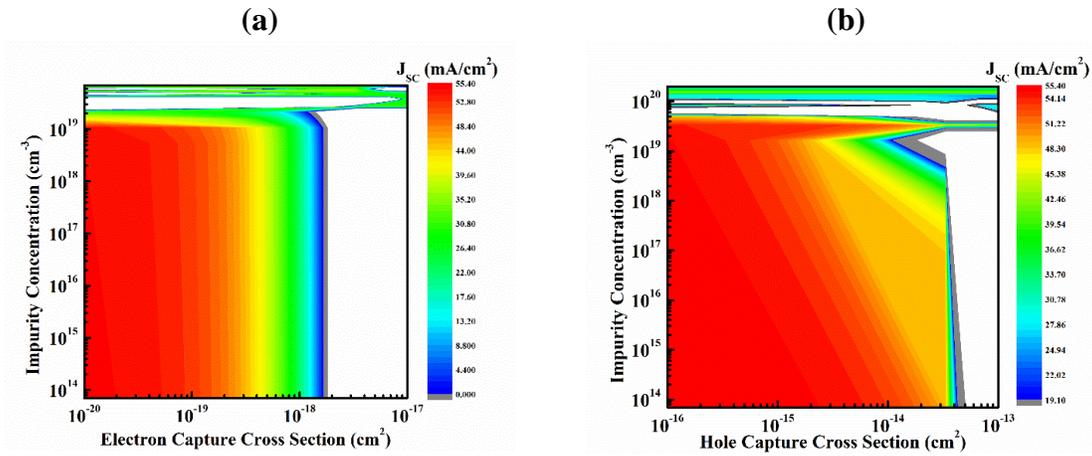

**Figure 10.** The variation of $J_{SC}$ with (a) impurity concentration and electron capture cross section and (b) impurity concentration and hole capture cross section of Zn impurity in Silicon.

However, it should be noted that the simulation of the IPV effect multilevel impurities as discussed above was performed without considering the band-to-band radiative and Auger recombination in the silicon that provides an optimized PCE of 35.4% with $V_{OC}$=0.789 V, $J_{SC}$=55.32 mA/cm², FF=80.99%, respectively. On the other hand, the consideration of radiative coefficient of $9.5 \times 10^{-15}$ cm³/s and Auger coefficients for electron and hole of $9.9 \times 10^{-32}$ and $2.8 \times 10^{-31}$ cm⁶/s, respectively[43] in the simulation yields a PCE of 34.97% with $V_{OC}$=0.773 V, $J_{SC}$=54.99 mA/cm², FF=82.26%, respectively.

### 3.3 Effect of multilevel impurities on quantum efficiency

The spectral response provides an assessment of how photons of various frequencies contribute to the short circuit current in the solar cell. For further confirmation of the fact that the additional contribution to the short circuit current comes due to addition of impurity levels as well as combination of suitable anode and cathode material, we perform numerical analysis of quantum efficiency (QE) as a function of spectral wavelength from 300 nm to 1800 nm for different impurity levels in non-flat band $p^+$-$n$-$n^+$ solar cell structure. The results are depicted in Fig. 11. In the figure, it is observed that the QE is almost zero at 1180 nm for the pristine solar cell structure. On the other hand, contribution of Tl impurity i.e. single level of IPV provides QE between 1200 to 1500 nm. Addition of further two impurity levels from the Zn provides 46% quantum efficiency on average in the range of 1500 nm to 1800 nm. The inclusion of different impurity levels increases the probability of absorbing sub-band gap photons and contributing more in current density as described in section 2.2 and 3.1.4. However, this type of higher longer wavelength light absorption has also been reported for Si and chalcogenide solar cells in which sub-band gap photons are absorbed by band tail-states-assisted (TSA) two-step photon upconversion resulting higher PCE of the solar cells[40,44-46].

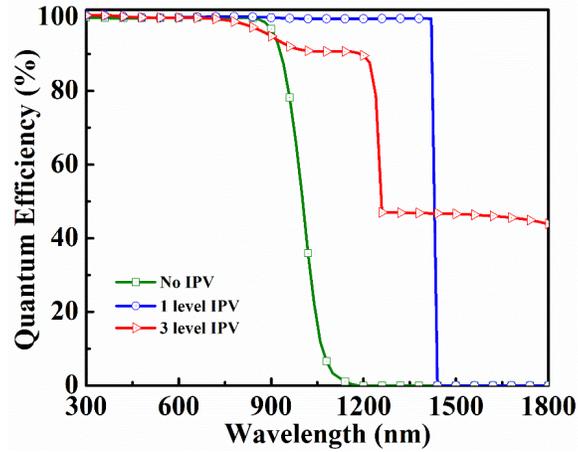

**Figure 11.** The QE of Si solar cells with multilevel impurities under non-flat band condition.

## 4. Conclusion

In this study, we perform numerical analysis of the IPV phenomena introducing three level of impurities with Tl and Zn using SCAPS-1D simulator. With an optimized $p^+$-$n$-$n^+$ structure with absorber thickness of 100 μm, doping concentration of $2\times10^{17}$ cm$^{-3}$, uniform impurity concentration of $7\times10^{13}$ cm$^{-3}$, the proposed solar cell resulted a short circuit current density of 55.32 mA/cm$^2$ and PCE of 35.4% with an open circuit voltage of 0.79 V. The improvement in short circuit current not only comes from the absorption of sub-band gap photons at different energy levels, but also from the addition of suitable metals Ni and Ta, respectively. To ensure the high current, thermal capture cross section of Zn has been varied within the reported ranges and an optimized value of capture cross sections have been found to be in the order of $10^{-20}$ and $10^{-16}$ cm$^2$ for electron and hole, respectively. While the IPV effect of Tl, expands spectral responds to 1100-1500 nm, the non-flat band combination as well as inclusion of Zn impurity results extended responses in the range of 1500-1800 nm. Therefore, it can be concluded that both IPV effect and role of proper metal work functions enhance current by sub-band gap absorption of longer wavelength photons.

## Acknowledgements

The authors highly appreciate Dr. Marc Burgelman, University of Gent, Belgium, for providing SCAPS simulation software.

**Corresponding authors:**

*E-mail: jak_apee@ru.ac.bd (Jaker Hossain).

**Notes:** The authors declare no competing financial interest.

*Appl. Mater. Sci.* **209**, 1002-1006 (2012).

18. Kasai, H., Sato, T. & Matsumura, H. Impurity photovoltaic effect in crystalline silicon solar cells. in *Conference Record of the IEEE Photovoltaic Specialists Conference* 215-218 (1997).

19. Schmeits, M. & Mani, A. A. Impurity photovoltaic effect in c-Si solar cells. A numerical study. *J. Appl. Phys.* **85**, 2207-2212 (1999).

20. Sze, S. M. & Ng, K. K. *Physics of Semiconductor Devices*. (2006).

21. Verschraegen, J., Khelifi, S., Burgelman, M. & Belghachi, A. Numerical Modeling of the Impurity Photovoltaic Effect ( IPV ) in SCAPS. in *21st European Photovoltaic Solar Energy Conference* 396–399 (2006).

22. Nevin, J. H. & Henderson, H. T. Thallium-doped silicon ionization and excitation levels by infrared absorption. *J. Appl. Phys.* **46**, 2130-2133(1975).

23. Yuan, J. *et al*. Impurity photovoltaic effect in silicon solar cells doped with two impurities. *Opt. Quantum Electron.* **46**, 1457-1465 (2014).

24. Schibli, E. & Milnes, A. G. Deep impurities in silicon. *Materials Science and Engineering* **2**, 173-180 (1967).

25. Yuan, J. *et al*. Photoemission cross section: A critical parameter in the impurity photovoltaic effect. *Chinese Phys. B* **26**, 018503 (2017).

26. Kumar, A. & Thakur, A. D. Impurity photovoltaic and split spectrum for efficiency gain in $Cu_2ZnSnS_4$ solar cells. *Optik* **238**, 166783 (2021).

27. Kumar, A. & Thakur, A. D. Role of contact work function, back surface field, and conduction band offset in $Cu_2ZnSnS_4$ solar cell. in *Jpn. J. Appl. Phys.* **57,** 08RC05 (2018).

28. Sklensky, A. F. & Bube, R. H. Photoelectronic properties of zinc impurity in silicon. *Phys. Rev. B* **6**, 1328-1336 (1972).